\documentclass{cernyrep}
\usepackage{texnames}
\usepackage[T1]{fontenc}
\usepackage{subfigure}
\usepackage{color}
\usepackage{amssymb}
\newcounter{query}

\begin{document}
\title{Fixed-Field Alternating-Gradient Accelerators}

\author{S.L. Sheehy}

\institute{University of Oxford, UK}

\maketitle 

\begin{abstract}
These notes provide an overview of Fixed-Field Alternating-Gradient (FFAG) accelerators for medical applications. We begin with a review of the basic principles of this type of accelerator, including the scaling and non-scaling types, highlighting beam dynamics issues that are of relevance to hadron accelerators. The potential of FFAG accelerators in the field of hadron therapy is discussed in detail, including an overview of existing medical FFAG designs. The options for FFAG treatment gantries are also considered.
\end{abstract}

\section{Preface}

These notes are broken down into two main parts. The first gives a general introduction to Fixed-Field Alternating-Gradient (FFAG) accelerators (often referred to simply as FFAGs) including an overview of the basic transverse dynamics of both the so-called `scaling' and the `non-scaling' variety, and a brief comparison with other machines. In the second part, I discuss the motivations for considering FFAGs for medical applications, focusing on charged-particle therapy (hadron therapy). I then introduce a number of FFAG designs specifically aimed at medical applications, including developments of FFAG gantries for medical use.

The FFAG accelerator is a class of circular accelerator that combines properties of both the cyclotron and the synchrotron. It uses a magnetic field which is constant in time, hence the `fixed-field', together with an increased focusing strength achieved using the `alternating-gradient' principle~\cite{bib:courantsnyder}. The RF acceleration scheme is usually variable-frequency, but in some specific instances a fixed-frequency system is possible.

Many accelerator physicists have difficulty when first encountering FFAGs, as they may have learned about only one example of an FFAG machine and therefore approach the topic with some preconceptions. Many questions arise, such as `aren't the magnets very large?', `have any of them actually been built?', and `isn't the FFAG only good for large muon beams?' The difficulty with these questions is that the modern FFAG is not really a single type of machine. This would be like assuming that a synchrotron can only be a light source, rather than a hadron collider or a medical accelerator. For the record, the answers to the questions are `sometimes', `yes', and `not only'. Of course, for synchrotrons it is easy to see that the application can change the layout, design, and properties of an accelerator. This is perhaps even more true in the case of the FFAG.

Starting with the idea that FFAGs are just accelerators which have both a fixed field and alternating-gradient focusing produces a large spectrum of designs. Some designs may have purely linear (quadrupole) magnets and fixed-frequency RF acceleration, whereas others have large-aperture magnets which produce a complex variation of the field with radius and a variable-frequency RF system. Most FFAGs have a very large dynamic aperture. This flexibility of FFAG design has only emerged in roughly the last 15 years and the field continues to be a rich source of novel developments.

From these lecture notes you should be able to answer questions like those above, and describe the various types of FFAG accelerators and some of the design principles behind them. I hope you will also learn of the potential and some of the challenges in utilizing these machines in medical applications.

\section{Fixed-field alternating-gradient accelerators}

\subsection{A historical perspective}
The concept of an FFAG accelerator is not new. This type of accelerator was invented in the 1950s and 1960s at the same time as the synchrotron was being developed. Much of the early work in developing FFAGs was carried out at the Midwestern Universities Research Association (MURA), but only electron FFAG accelerators were constructed at the time~\cite{bib:jones09}. It is interesting also to note that in Soviet terminology the FFAG is known as a `ring phasotron'~\cite{bib:kolomensky}. It was not until the 1990s that interest in this type of accelerator re-emerged in Japan, with a particular focus on what such machines could offer as hadron accelerators. For a little more on the history, see Ref.~\cite{bib:machida2012} and the references therein.

A particular area of recent interest in the field of FFAGs is their potential for high-intensity operation, because of their high repetition rate, large acceptance, simpler and cheaper power supplies, and flexibility of the RF acceleration system. High intensity may be required for some medical applications  (such as radioisotope production), but for hadron therapy the beam currents are typically low. For this reason, we shall not discuss high intensities here but instead focus on the advantages in terms of high repetition rate, flexibility, and simple magnet power supplies. The impact of these qualities will be discussed in Section~\ref{section4}.

\subsection{The original or `scaling' FFAG}
In 1943 Marcus Oliphant described the idea of the synchrotron as follows:
\begin{quotation}
Particles should be constrained to move in a circle of constant radius thus enabling the use of an annular ring of magnetic field \ldots which would be varied in such a way that the radius of curvature remains constant as the particles gain energy through successive accelerations.
\end{quotation}
He intended that the magnetic field should be varied temporally and the beam should always follow the same annulus~\cite{bib:oliphant}. However, in principle there is no reason why the annulus may not change radius and the field vary spatially rather than temporally. This is the fundamental idea behind the FFAG. A large variation of the field with radius will constrain the change in radius of the orbits; this can lead to a larger field increase with radius and more compact orbits than in a cyclotron. This is the original type of FFAG, which we now call `scaling'.

To introduce the scaling FFAG, it is useful to review the four main different types of circular accelerator, which we can classify by the magnetic field they use to guide the particles~\cite{bib:kolomensky}. These are the following.
\begin{enumerate}
\item{Fixed-field constant-gradient accelerators, including conventional cyclotrons, synchrocyclotrons, and microtrons.}
\item{Pulsed-field constant-gradient accelerators, which includes weak-focusing synchrotrons and betatrons.}
\item{Pulsed-field alternating-gradient accelerators, which are the well-known AG synchrotrons.}
\item{Fixed-field alternating-gradient accelerator, otherwise known as FFAGs.}
\end{enumerate}

The fourth variety, FFAG accelerators, were proposed independently in the early 1950s by Ohkawa in Japan~\cite{bib:ohkawa}, Symon \emph{et al.} in the United States~\cite{bib:symonkerst}, and Kolomensky in Russia~\cite{bib:kolomensky57}. Symon \emph{et al.}  proposed:
\begin{quotation}
A type of circular accelerator with magnetic guide fields which are constant in time, and which can accommodate stable orbits at all energies from injection to output energy.
\end{quotation}
This relies on introducing sectors with a reversed magnetic field into a cyclotron-like machine, producing strong focusing throughout the energy range. The field may rise rapidly with radius such that the orbits are relatively compact over a large energy range. It is possible to accelerate both light particles (electrons) and heavier particles (hadrons) to relativistic energies with this method. The time-independent magnetic field means that the repetition rate can be much higher than in a pulsed-field alternating-gradient machine (i.e., synchrotron), as the RF modulation can be on a much shorter time-scale than the modulation of a magnetic field. The consequences of this for medical applications will be discussed later.

The field is arranged in such a way that the increase in gradient with momentum results in the beam experiencing the same focusing independent of radius. This means that the betatron tunes are constant for all orbits. This constant focusing (or constant betatron tune) is ensured if two conditions are met. First, the field index $k$ must be constant, where we can define $k$ in terms of the bending radius $\rho$, the vertical magnetic field $B_{y}$, and its derivative in the horizontal direction $x$:
\begin{equation}
k= -\frac{\rho}{B_{y}}\frac{\partial B_{y}}{\partial x} .
\label{eqn:kvalue}
\end{equation}
Therefore we require
\begin{equation}
\frac{\partial k}{\partial p}\bigg|_{\theta = \mathrm{const.}} = 0 .
\label{eqn:cardinal1}
\end{equation}

The second requirement is that the shape of the particle orbits remains constant as the size of the orbits `scales' with energy, such that each higher-energy orbit is a geometrically similar enlargement of the lower-energy orbits as described by the following equation, derived by Kolomensky~\cite{bib:kolomensky}:
\begin{equation}
\frac{\partial}{\partial p}\left(\frac{\rho_{0}}{\rho}\right)\bigg|_{\theta = \mathrm{const.}} = 0.
\label{eqn:cardinal2}
\end{equation}
If the field meets these two conditions, the FFAG is referred to as being of the `scaling' variety.

\begin{figure}[h!]
\centering\includegraphics[width=.5\linewidth]{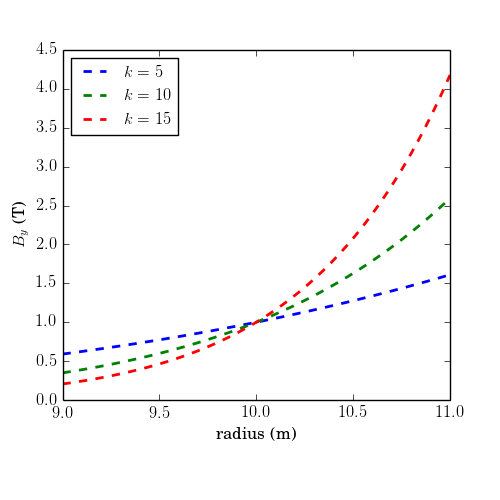}
\caption{An example of the characteristic scaling law for different values of the field index $k$, with $r_{0}=10$~m and $B_{0}=1.0$~T.}
\label{fig:scalinglaw}
\end{figure}

To satisfy these requirements, we use a magnetic field that increases with radius. The particular shape of the field is given by the $r^{k}$ law of the following equation, which describes the increase in field with radius $r$ with respect to a reference radius $r_{0}$, where the field increase is characterized by the field index, $k$:
\begin{equation}
B_{y} = B_{0} \left (\frac{r}{r_{0}} \right)^{k} .
\label{eqn:scalinglaw}
\end{equation}
The field is shown in Fig.~\ref{fig:scalinglaw}. It should be clear that for a given value of the field index $k$, the field at a smaller radius not is only lower but also has a lower gradient. As the momentum increases and particles move to higher radius, the value of the confining field increases along with the field gradient. Of course, this is the field for only one of the two alternating-gradient types, the `F' or focusing type. The field for the `D' (defocusing) type has the opposite sign, as expected.

\begin{figure}[h!]
\centering\includegraphics[width=.5\linewidth]{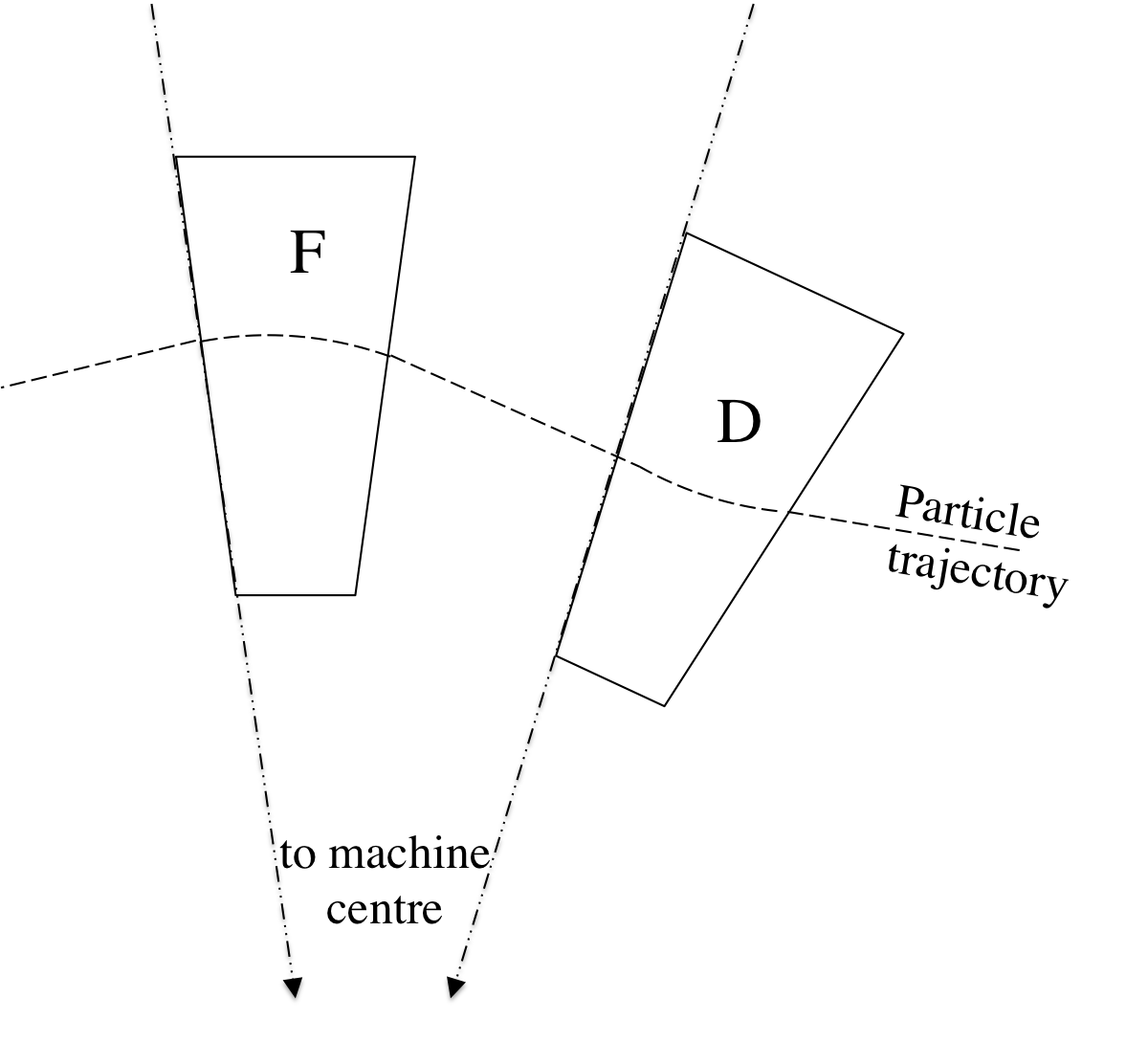}
\caption{Radial-sector FFAG layout with reverse bend}
\label{fig:scalingffag}
\end{figure}

The fact that the different sectors in the scaling FFAG have reverse-polarity magnetic fields means that the length of the orbit and the mean radius of the machine are necessarily larger than if there were no reverse fields, as shown in Fig.~\ref{fig:scalingffag}. This is necessary to ensure stability in both planes and is the main disadvantage of the scaling FFAG. This can be partly or fully overcome, however, as the maximum magnetic-field value can be higher than that in a synchrotron with a time-varying field, which can help to constrain the machine size. In addition, there are no separate-function magnets, so provided sufficient space is left for injection, extraction, accelerating cavities, vacuum ports, and so forth, the machine may be made compact as the main magnets perform all the required functions simultaneously. The FFAG layout in Fig.~\ref{fig:scalingffag} is called a `radial-sector' layout as the faces of the magnets lie along radial lines from the machine centre.

\begin{figure}[h!]
\centering\includegraphics[width=.5\linewidth]{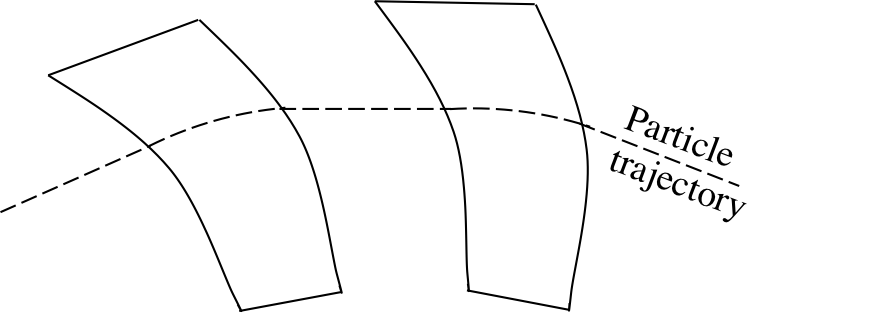}
\caption{Spiral FFAG layout, removing the need for reverse bends}
\label{fig:spiralffag}
\end{figure}

There is in fact a second type of FFAG which does away with the need for the reverse-polarity field, known as the `spiral' FFAG. This is formed from a succession of hills and valleys of field distributed in the azimuthal direction, where the magnets have a spiral angle with respect to the beam as shown in Fig.~\ref{fig:spiralffag}. In this case the beam does not enter the magnet exactly perpendicular to the face of the magnet, and thus edge focusing results. This removes the need for the opposing reverse bending magnet, while maintaining strong focusing. In the spiral FFAG, the focusing can also be made independent of momentum.

The field in a spiral FFAG in general has the form
\begin{equation}
B(r,\theta) = B_{0} \left( \frac{r}{r_{0}} \right)^{k} F(\vartheta) ,
\label{eqn:scalinglaw2}
\end{equation}
where we call $\vartheta$ a generalized angle. This is related to the usual azimuth by
\begin{equation}
\vartheta = \theta - \tan\zeta \ln\frac{r}{r_{0}}.
\end{equation}
The value of $\zeta$ is the angle between the field spiral and the radial direction, and $\tan\zeta$ is constant. As before, $B_{0}$ and $r_{0}$ are constants. The function $F(\vartheta)$ is an arbitrary periodic function with period $\vartheta_{0}=2\pi/N$, and $N$ is the number of periods. In the radial-sector scaling FFAG that we discussed earlier, all the orbits enlarge and remain geometrically similar, but in the spiral FFAG they also turn about the centre. This can be seen in an example spiral FFAG design for the RACCAM project in Section~\ref{section4}.

In terms of beam dynamics, it is useful to compare the scaling FFAG with a synchrotron. Modern synchrotrons employ the principle of alternating-gradient or `strong' focusing~\cite{bib:courantlivingston52,bib:courantsnyder58}, in which alternating focusing and defocusing magnets lead to much stronger focusing forces in the transverse plane than in constant-gradient weak-focusing synchrotrons. This alternating-gradient focusing is also employed in the FFAG. The transverse beam dynamics in the FFAG is therefore much the same as in the synchrotron, at least for a single orbit or energy, in the sense that we may discuss beta functions, dispersion, and so forth. The difference is that in this case the field is highly nonlinear and these transverse optics functions may vary with radius.

\section{The non-scaling FFAG}

The non-scaling FFAG allows the strict scaling laws applied in the original scaling FFAG to be relaxed. The idea of violating the strict scaling law of the FFAG occurred to Kent Terwilliger and Lawrence W. Jones in the 1950s~\cite{bib:jones91}, but such a machine was never pursued. Two of the main disadvantages of the original FFAG are the highly nonlinear magnetic field required and the large aperture of the magnets and RF arising from the shift of the orbit with energy, which can be up to the order of 0.5--1.0~m. The non-scaling FFAG arose from the question ``what if we violate the scaling law?'' Or, more specifically, ``What if we take a line tangent to the scaling law in Fig.~\ref{fig:scalinglaw}, such that the field is linear with radius?'' This radical idea led to the linear non-scaling FFAG and was proposed in the 1990s~\cite{bib:johnstone99}.

\subsection{Linear non-scaling FFAG}

The linear non-scaling FFAG is so called because it uses only up to linear focusing elements, that is, quadrupole and dipole fields. When only quadrupoles and no higher-order multipoles are used for focusing, the beam shifts outward with acceleration because of dispersion and sees a reduced level of focusing. This is really like considering a synchrotron where we do not ramp the magnets with time. In the scaling FFAG, we got around this by varying the gradient with the momentum and by making the beam pipe wider to allow for the orbits moving. But in the linear non-scaling FFAG we ignore the scaling law and any focusing issues for now, which allows us to increase the gradient and reduce the dispersion function even further to reduce the shift of the orbit with momentum. To achieve this, a linear non-scaling FFAG lattice may use normal bending with a defocusing `D' quadrupole and reverse bending with a focusing `F' quadrupole, and may (or may not, depending on the design) change at high momentum to use the `D' quadrupole for reverse bending and the `F' for normal bending, as described in Ref.~\cite{bib:machida2012}.

One must then ask what happens to the beam dynamics in such an accelerator. One consequence is that the orbits no longer `scale', so they are no longer geometrically similar at different energies. The orbits can be made much more compact than in the scaling FFAG. However, the most dramatic difference is that the betatron tunes are no longer constant with energy. This may seem surprising if you have worked on (almost) any other type of accelerator, as the betatron tunes are usually designed to be kept constant. In the linear non-scaling FFAG, they vary dramatically throughout the acceleration cycle, crossing not just high-order betatron resonances but also integer resonances.

In theory, if the acceleration is fast enough, the beam may be able to cross betatron resonances before they have time to build up, and therefore any amplitude growth effects may be mitigated. How fast this crossing needs to be depends on imperfections and alignment errors in the machine, and clearly necessitates a fast acceleration rate. In fact, the linear non-scaling FFAG was proposed in the context of muon acceleration, where very fast acceleration before the muons decay is an absolute requirement. The many questions surrounding the dynamics of such a machine led to the construction of the first non-scaling FFAG accelerator, known as EMMA.

\subsection{The EMMA non-scaling FFAG}

\begin{figure}[h!]
\centering\includegraphics[width=.5\linewidth]{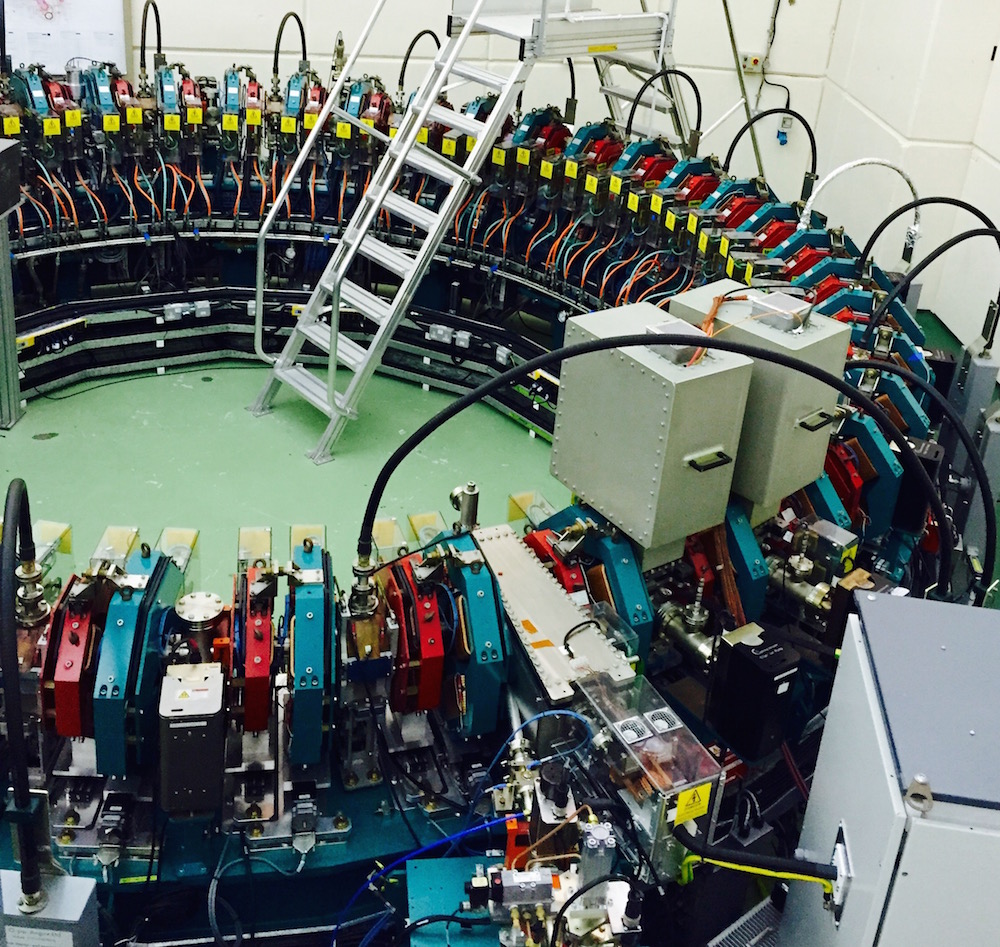}
\caption{The EMMA non-scaling FFAG at the STFC Daresbury Laboratory, UK}
\label{fig:emmaring}
\end{figure}

EMMA is the Electron Machine for Many Applications, a 10--20~MeV electron accelerator constructed at Daresbury Laboratory to demonstrate the technology of the linear non-scaling FFAG accelerator. The accelerator is shown in Fig.~\ref{fig:emmaring}, and measures just over 5~m in diameter. Despite the name, EMMA is in fact a demonstrator machine; the primary purpose of its commissioning in 2010--2011 was to investigate some of the key dynamics issues and technical issues involved in realizing the technology. This being demonstrated, it aimed to show that the non-scaling FFAG was a viable technology for use in `many applications' ranging from medical use and muon acceleration to high-power proton accelerators.

A few key points of interest, of relevance to medical accelerators, were learned from the EMMA experiment.\footnote{EMMA also demonstrated a number of other interesting accelerator physics concepts, including a novel acceleration mechanism known as `serpentine acceleration'.} Most importantly, EMMA demonstrated the successful operation of a linear non-scaling FFAG, bringing this technology into reality. Next, EMMA demonstrated that fast crossing of integer betatron resonances is possible if the beam is accelerated quickly enough. However, EMMA accelerates in just 10 turns from injection to extraction and it is not practicable to accelerate a hadron beam so quickly over the energies required for medical use. We shall return to this point in our discussion of hadron FFAG design studies. More information about EMMA can be found in Refs.~\cite{bib:emmanature, bib:barlow10}.

\section{FFAGs for hadron therapy}
\label{section4}

In this section we turn to addressing the specific idea of FFAG accelerators for medical applications, in particular hadron therapy. State-of-the-art hadron therapy (proton and light-ion therapy) centres must provide beams for patient treatment with unprecedented precision, flexibility, and reliability within a hospital environment. This can pose challenges for present accelerator technology, in terms of both beam requirements and considerations of machine size and cost. Existing technological options may meet present requirements, but one should ask whether they are the optimal solution for the future of this treatment modality. Can we do better?

In accelerator terms, the beam energies required for therapy are relatively low compared with the multi-TeV energy range required for high-energy physics experiments. Proton energies up to 250~MeV are required for treatment, and up to 330~MeV if online proton radiography is incorporated. For heavier ions, full-body-treatment-energy C$^{6+}$ is equivalent in magnetic rigidity to protons with an energy of around 1.2~GeV. These energies can be achieved with a number of different types of accelerator, including synchrotrons, cyclotrons, and linear accelerators. Because of the limited space available in a clinical environment, a circular accelerator is usually chosen as the main accelerator for a facility.\footnote{However, recent developments such as the `Cyclinac' design are being studied for this application~\cite{bib:amaldi2009}.} There are a number of possible choices of circular accelerator. Most existing or planned facilities use the established technology of either the synchrotron or the cyclotron.

Cyclotrons have a limited energy range, as it is difficult to maintain vertical focusing at high energy and the magnets become increasingly unwieldy as the energy and thus machine size grow. They are also limited by their fixed extraction energy. For hadron therapy, this means in practice that the beam energy must be adjusted between the accelerator and the patient using an energy degrader. Although a cyclotron can deliver ample dose to the patient, the use of energy degraders and passive scattering systems has led to concerns about activation and radiation protection~\cite{bib:pedroni00}. State-of-the-art systems ought to use active scanning systems and active energy variation to deliver the best-quality treatment.

Synchrotrons operate with a pulsed beam; however, most synchrotrons have a slow cycle rate, on the order of 1~Hz. Rapid-cycling synchrotrons have been proposed for this application, with rates up to around 50~Hz. Their main advantage is their flexibility, particularly in terms of energy reach and easy variable-energy extraction. However, the pulsed nature of these machines makes scanning rather slow. This problem has been partly overcome by using a stable slow extraction of the beam, although the energy variability is still limited by the repetition rate of the synchrotron. In addition, synchrotrons are generally much larger than cyclotrons.

This discussion leads us to ask where the FFAG accelerator might enter the picture. With a fixed magnetic field, the repetition or cycle rate can be much higher than in a synchrotron, up to the range of kHz, limited only by the speed of the RF system. With variable-energy extraction, this high repetition rate could enable slice-to-slice energy variation without limiting the dose rate for full 3D conformal irradiation, or even 3D tumour motion tracking during treatment. Such a machine is not limited in energy range by its dynamics like a cyclotron, and so can easily reach the energies required for therapy. If fast extraction can be achieved at any energy, one may even conceive of a machine which operates with different ion species on a pulse-by-pulse basis, using some pulses for imaging and some for treatment while using the same acceleration and beam delivery system. Of course, this requires that the extraction line, beamlines, gantry, diagnostics, and quality assurance are able to handle such flexibility. In essence, the FFAG has the potential to remove some of the limitations of existing technologies and provide a flexible, relatively compact alternative.

\subsection{The KEK 150~MeV FFAG}

\begin{figure}
\centering\includegraphics[width=.5\linewidth]{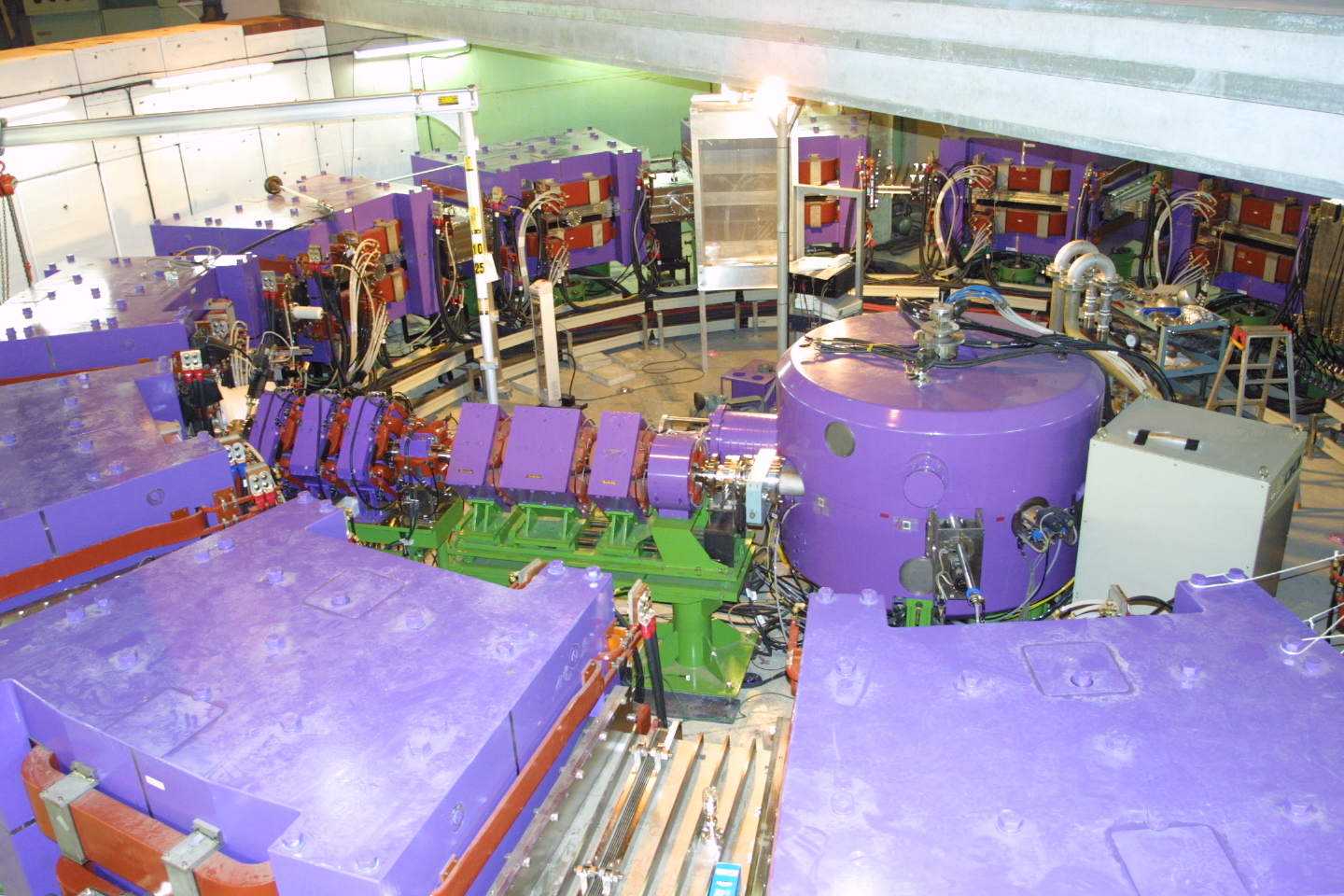}
\caption{The KEK 150~MeV scaling FFAG, Japan}
\label{fig:kekring}
\end{figure}

The first prototype FFAG for medical applications was a 150~MeV proton accelerator at KEK in Japan. This was one of the first hadron FFAGs ever to be built and is a radial-sector scaling FFAG using a 12-sector DFD triplet lattice structure. The orbits vary from an average radius of 4.47~m at 12~MeV to 5.2~m at 150~MeV at the centre of the F magnet. This machine was as much about proof of technology as it was about suitability for medical applications. A number of key innovations were made which addressed some of the technology challenges of the scaling FFAG, including wide-aperture magnetic-alloy RF cavities for acceleration~\cite{bib:yonemura} and the development of a `return-yoke-free' main magnet triplet, which eased beam injection and extraction by allowing traversal through the side of the main magnets. A very similar machine for studies of accelerator-driven systems is also in operation at Kyoto University Research Reactor Institute, and the original KEK machine is now in use at Kyushu University. The machine is shown in Fig.~\ref{fig:kekring} along with the cyclotron injector and transport line.

\subsection{The RACCAM project}

\begin{figure}[h!]
\subfigure[]{\includegraphics[width = .45\linewidth]{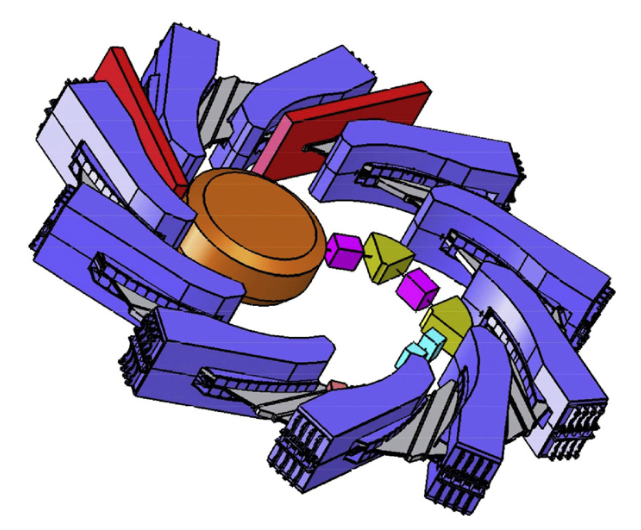}}
\subfigure[]{\includegraphics[width = .45\linewidth]{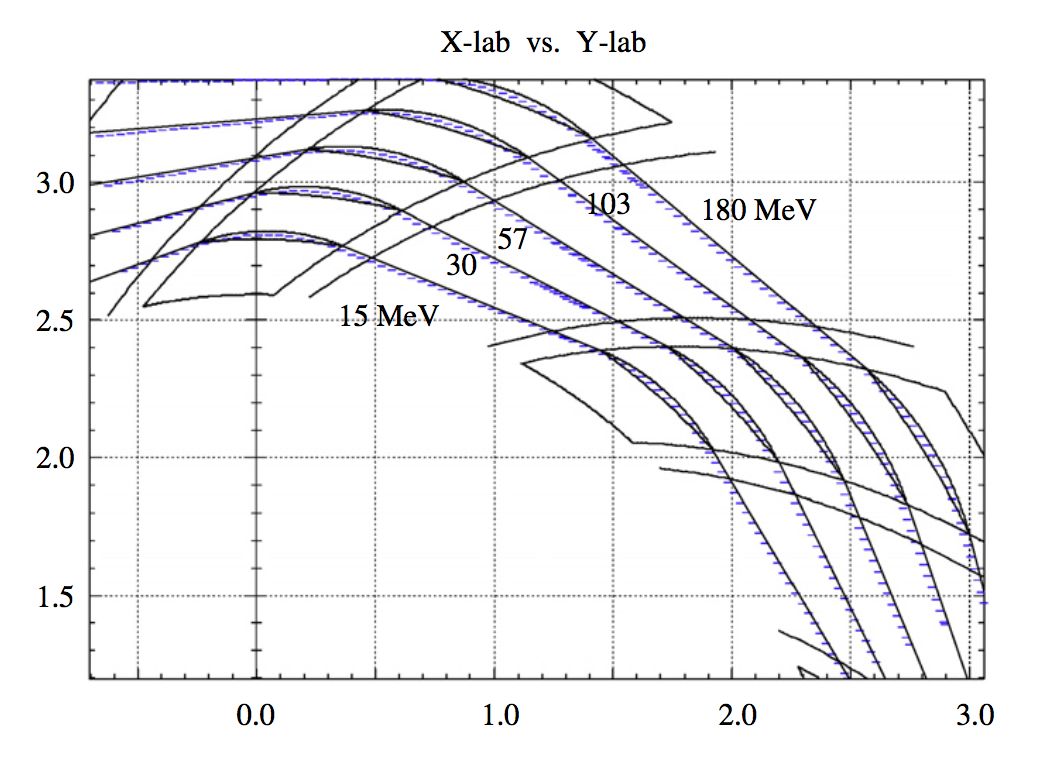}}
\caption{(a) Machine layout and (b) orbits in the RACCAM FFAG}
\label{fig:raccam}
\end{figure}

As discussed earlier, one issue with the radial-sector scaling FFAG is that it has a relatively large circumference because of the reverse bends, but the spiral FFAG does not have this issue. The RACCAM project, which ran from 2006 to 2008 in France, was a university and industry collaboration to design and prototype a spiral scaling FFAG for medical applications. It combined some key benefits, including variable-energy fast extraction and multiport extraction to different beamlines. The machine was designed to accelerate beams over a variable energy range from as low as 5.55~MeV to a maximum of 180~MeV with a repetition rate of at least 100~Hz. The design principles of the machine can be found in Ref.~\cite{bib:antoine09} and the layout and orbits for varying energies can be seen in Fig.~\ref{fig:raccam}.

As part of the RACCAM project, a suitable spiral FFAG magnet was designed in iteration with beam dynamics studies, and a prototype magnet was fabricated and measured. Details of the design can be found in Ref.~\cite{bib:planche09}. Along with the machine study and the technical design, a lot of work was undertaken in an attempt to optimize the facility itself. The collaboration carried out a study on how best to utilize multiple treatment rooms taking advantage of a machine with multiple extraction ports~\cite{bib:meot15}.

\subsection{Linear non-scaling FFAG designs}

Around the time the linear non-scaling FFAG concept was invented, a number of designs to apply this concept to hadron therapy arose. One example which was studied in detail was designed by Keil \emph{et al.}~\cite{bib:keil07} and consists of three concentric rings, to produce full-energy protons and carbon ions for hadron therapy.

\begin{figure}[h!]
\centering\includegraphics[width=.5\linewidth]{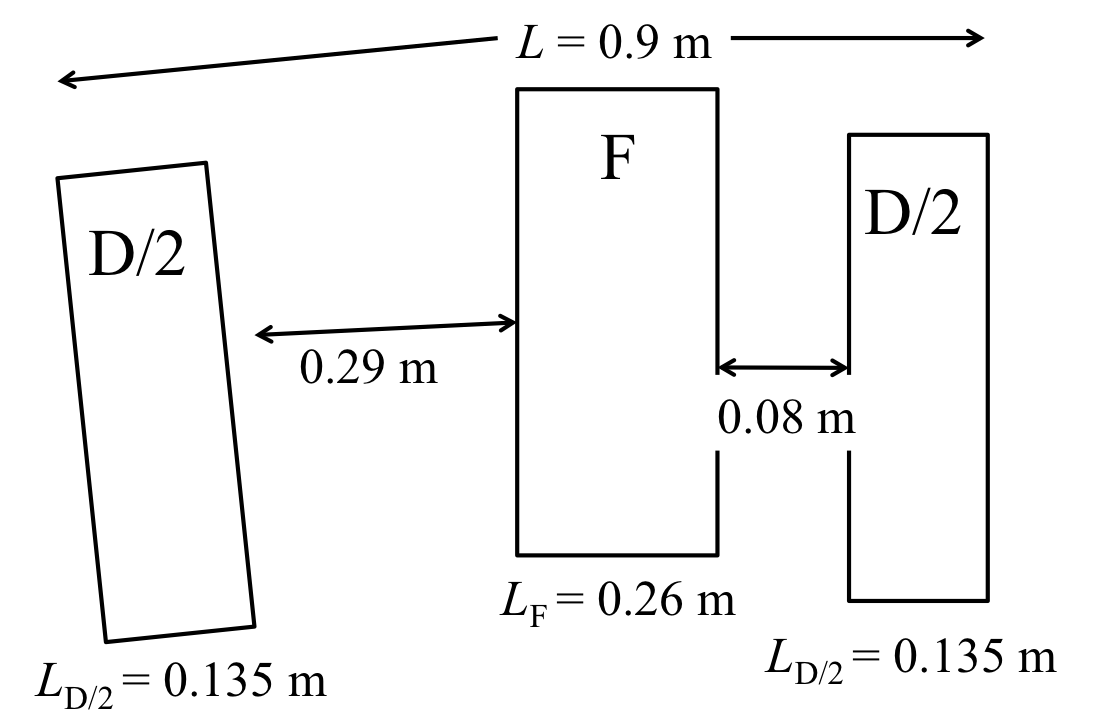}
\caption{Layout of one cell of the second (middle) ring from centre to centre of the D magnets in two adjacent cells}
\label{fig:kstcelllayout}
\end{figure}

\begin{figure}[h!]
\centering\includegraphics[width=.6\linewidth]{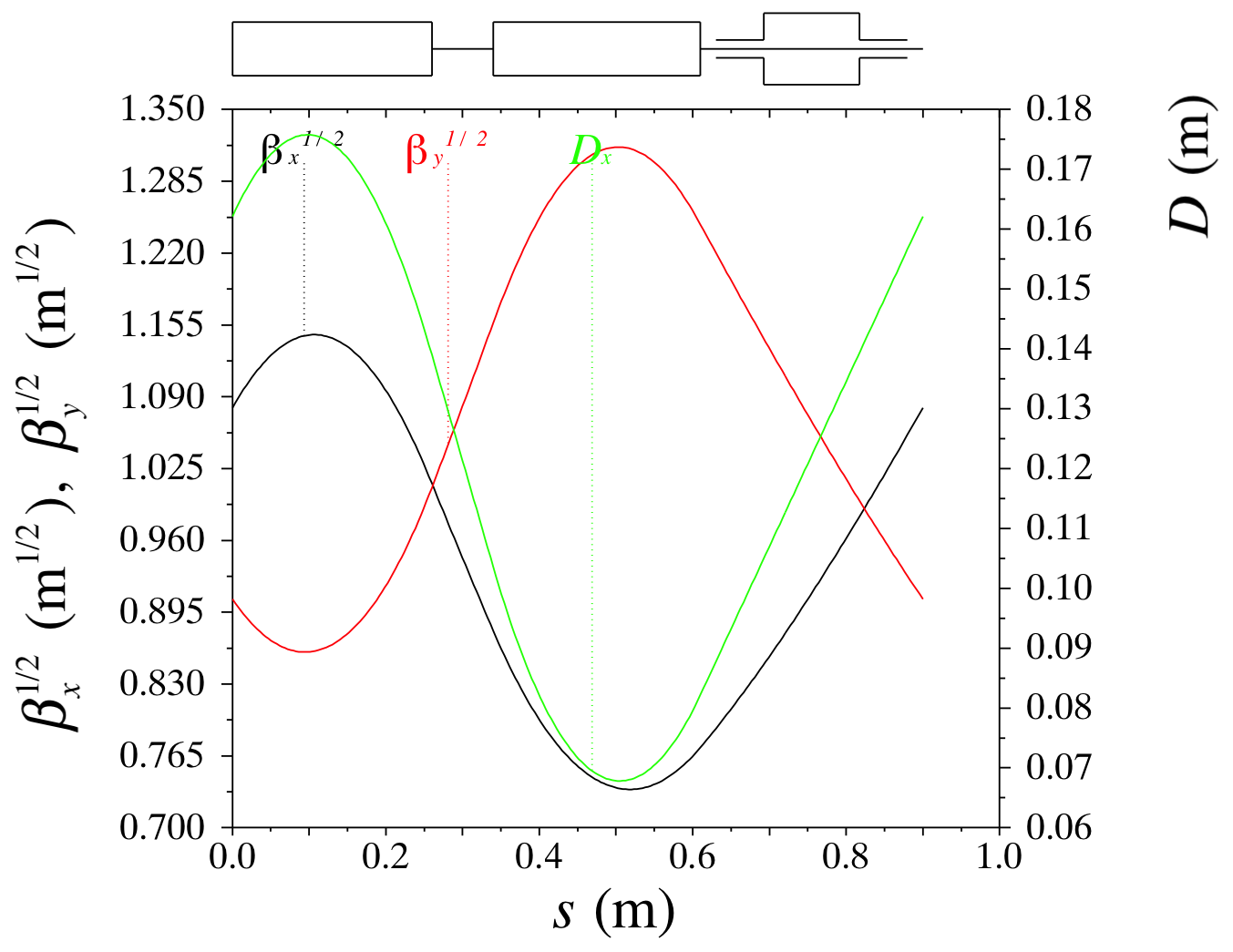}
\caption{Beta functions and dispersion in the Keil \emph{et al.} non-scaling FFAG for hadron therapy}
\label{fig:keilring2}
\end{figure}

This is the simplest type of non-scaling FFAG, the linear version, consisting of F and D quadrupoles, providing alternating-gradient (F/D) strong focusing. In this lattice, the use of F and D doublet magnets is proposed, in which a radial offset between the F and D quadrupoles provides the dipole bending field, removing the need for separate dipole magnets, as in the EMMA design. The lattice of the second ring, which covers the 31--250~MeV proton energy range (and the equivalent for carbon ions), comprises 48 FD doublet cells and the ring radius is around 6.9~m. The layout of the lattice cell can be seen in Fig.~\ref{fig:kstcelllayout} and the optics in Fig.~\ref{fig:keilring2}.

\begin{figure}
\centering\includegraphics[width=.7\linewidth]{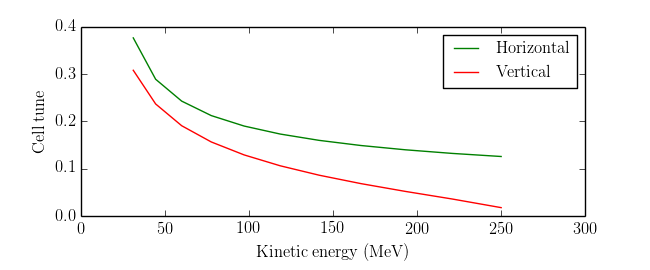}
\caption{Variation of betatron tune with acceleration. The cell tune is shown; the machine tune can be found by multiplying by the number of cells, which is 48 in this ring, resulting in the tunes crossing around 10 integers in each plane.}
\label{fig:tunes_kst}
\end{figure}

As we now know, the betatron tune in a linear non-scaling FFAG changes during acceleration; the variation of the tune in this lattice is shown in Fig.~\ref{fig:tunes_kst}. This changing tune will cross integer and half-integer betatron resonances, where any errors in the lattice will result in small kicks to the beam which will build up with each subsequent turn, potentially damaging the beam quality. Acceleration in this case occurs in roughly 1000 turns, which is far slower than the 10-turn acceleration in EMMA.

Detailed studies were carried out on the sensitivity of such a design to alignment errors, and a comparison was made with an alternative design by Trbojevic~\cite{bib:trbojevic08}. The results showed that alignment tolerances in such a machine may be as tight as 10 microns, an extremely stringent requirement. One may improve things by optimizing the lattice for the application so as to cross fewer resonances, but it is likely that resonance crossing will remain an issue~\cite{bib:sheehy10, bib:sheehythesis}. The topic of resonance crossing in non-scaling FFAGs is an ongoing area of active research~\cite{bib:moriya15}. In addition, one only needs to note the density of such a lattice to realize that injection and extraction and the lack of long straight sections may be problematic in an operational machine. However, the advantages of the linear non-scaling FFAG may be realized if the machine is made non-symmetric with long straight insertions. The principle may also be used to optimize the treatment gantry rather than the accelerator itself, an option which we shall discuss later.

\subsection{PAMELA: Particle Accelerator for MEdicaL Applications}

\begin{figure}
\centering\includegraphics[width=.5\linewidth]{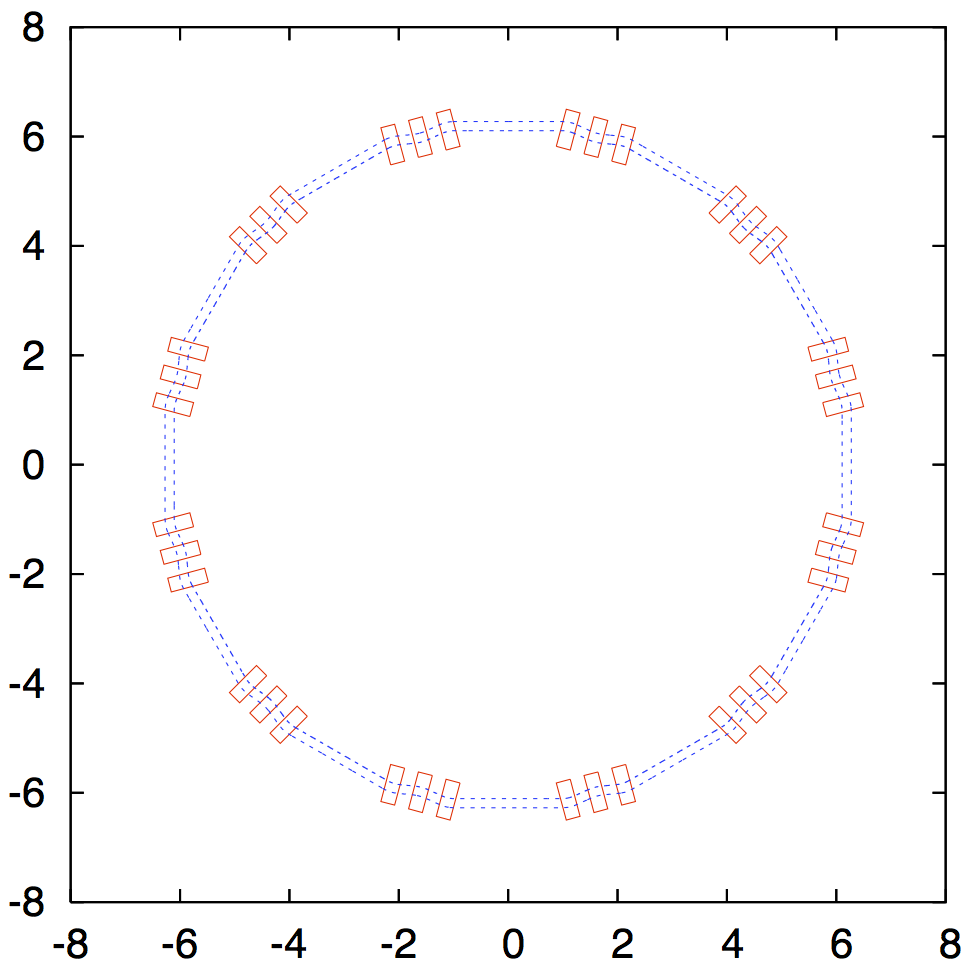}
\caption{Layout of the PAMELA proton ring lattice showing the low-energy (inner) and high-energy (outer) orbits. The axis units are metres.}
\label{fig:pamela1}
\end{figure}

As part of the CONFORM (COnstruction of a Non-scaling FFAG for Oncology, Research and Medicine) project, which included the construction of the EMMA non-scaling FFAG, a design study for a hadron non-scaling FFAG for hadron therapy was undertaken from 2007 to 2011. Known as the PAMELA project, this was a design study aimed at applying the concept of a non-scaling FFAG to hadron therapy. The design principle was based on two rings to cover the full energy range of both protons and carbon ions up to an equivalent magnetic rigidity of 6.7~T~m, that is, 440~MeV/nucleon for C$^{6+}$ ions (Fig.~\ref{fig:pamela1}). The machine was designed to operate with a very fast repetition rate of 1~kHz.

The starting point for the study was a linear non-scaling FFAG. However, the concern about beam deterioration from resonance crossing discussed in the previous section was significant enough to motivate studies of alternative designs of FFAGs for medical use. This resulted in a non-linear FFAG design. In each ring, the design strategy resulted in a variation of the total betatron tunes with acceleration which was well within half an integer (i.e., $\Delta\nu_\mathrm{total} = n_\mathrm{cell} \, \Delta\nu_\mathrm{cell} < 0.5$). It also maintained some advantageous non-scaling properties such as small orbit excursion and simpler magnets which are easy to align.

The steps of the design were as follows:
\begin{enumerate}
\item{The designers started with a scaling law based on a radial-sector FFAG with an FDF triplet lattice.}
\item{The second stable region of Hill's equation was used, with a horizontal phase advance per cell greater than 180$^{\degree}$. This allowed a larger field index to be used, resulting in a reduction in the orbit excursion by a factor of roughly five compared with the first stability region~\cite{bib:machida09}.}
\item{The field was decomposed into its multipole components and the expansion was truncated at some low order, usually octupole or decapole, to simplify the magnets.}
\item{The shape of the magnets was changed to rectangular rather than sector-shaped, and aligned on a straight line rather than along an arc to simplify construction.}
\item{The field profile was finally optimized such that the variation in betatron tune was minimized throughout acceleration.}
\end{enumerate}

\begin{figure}
\centering\includegraphics[width=.8\linewidth]{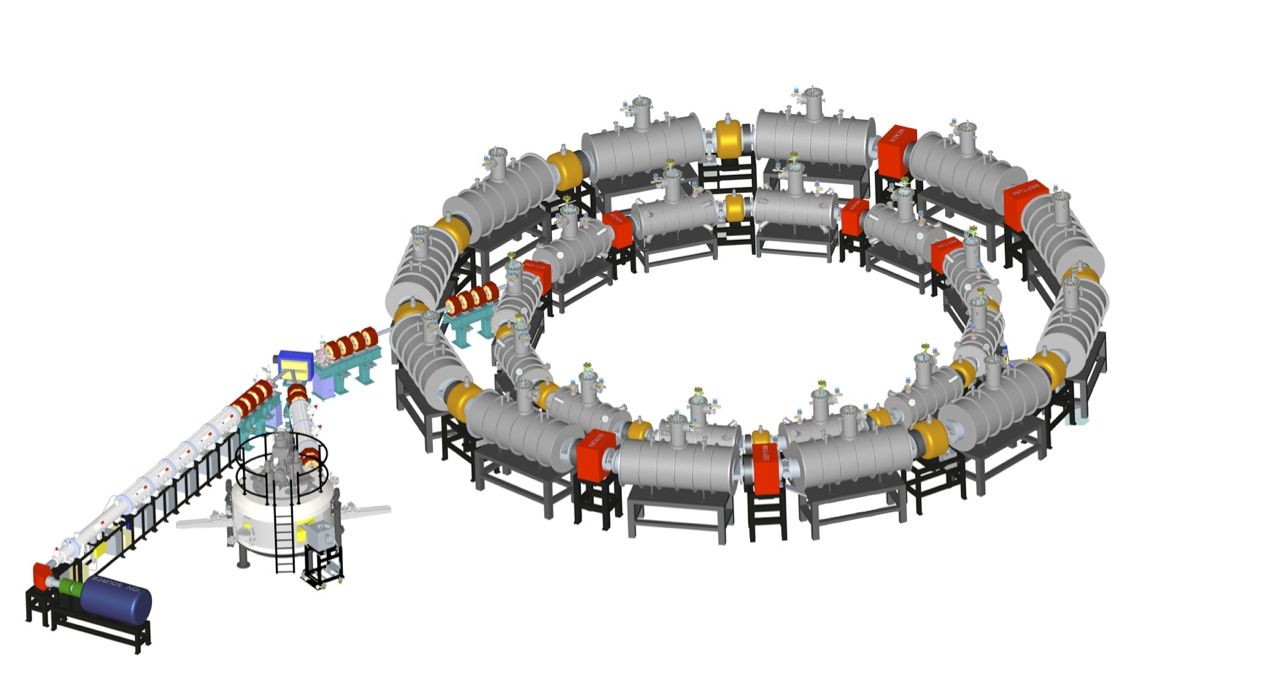}
\caption{Layout of the PAMELA facility}
\label{fig:pamelalayout}
\end{figure}

The resulting machine was studied up to engineering design level, including a novel superconducting magnet design~\cite{bib:witte12}; studies of RF cavities; engineering design of the cryostats, injectors, injection and extraction magnets, and beamlines; and a first look at a gantry system~\cite{bib:peach13}. The layout of the injectors and two main rings is shown in Fig.~\ref{fig:pamelalayout}.

\subsection{NORMA: NORmal conducting Medical Accelerator}

The NORMA study followed on from the PAMELA design to address some of the issues that arose during that study and to iterate the design. For example, it was felt that the PAMELA machine might have been closer to implementation if normal-conducting magnets had been used instead of the novel superconducting design. The NORMA design also extended the energy range of a single ring to provide beams up to 350~MeV which could travel through a patient to enable proton radiography. Although the design was for protons only (unlike PAMELA, which was for both protons and carbon ions), it provided detailed studies and optimization based on dynamic aperture~\cite{bib:garland15}.

\begin{figure}
\centering\includegraphics[width=.6\linewidth]{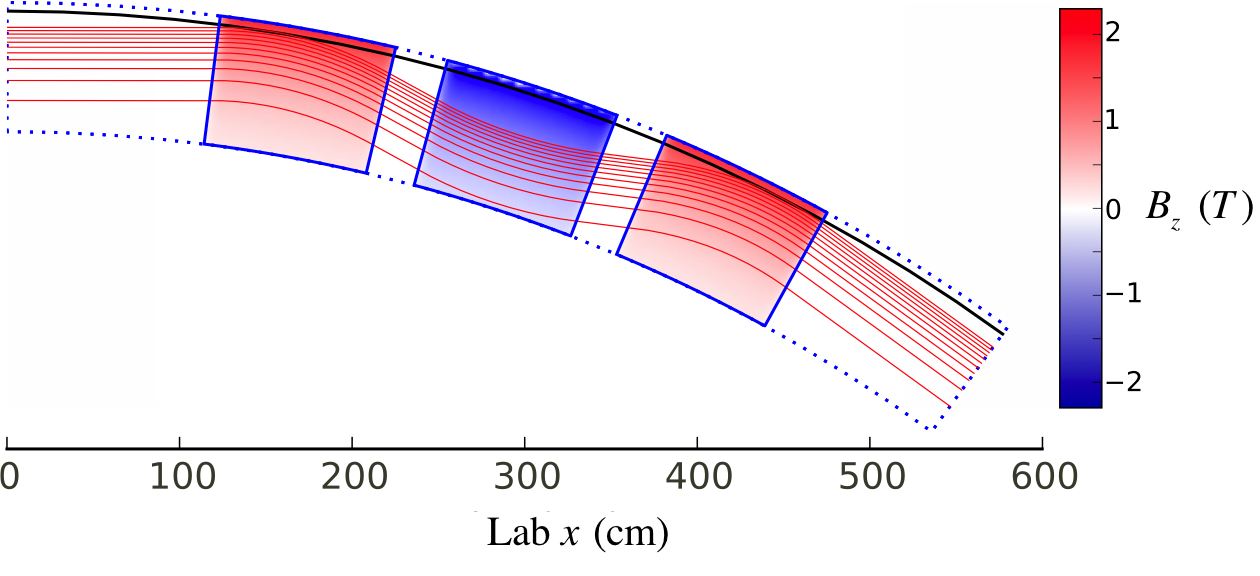}
\caption{Layout of the NORMA cell, showing the magnetic field throughout the cell and trajectories of low-energy (inner) to high-energy (outer) orbits.}
\label{fig:norma1}
\end{figure}

\begin{figure}
\centering\includegraphics[width=.6\linewidth]{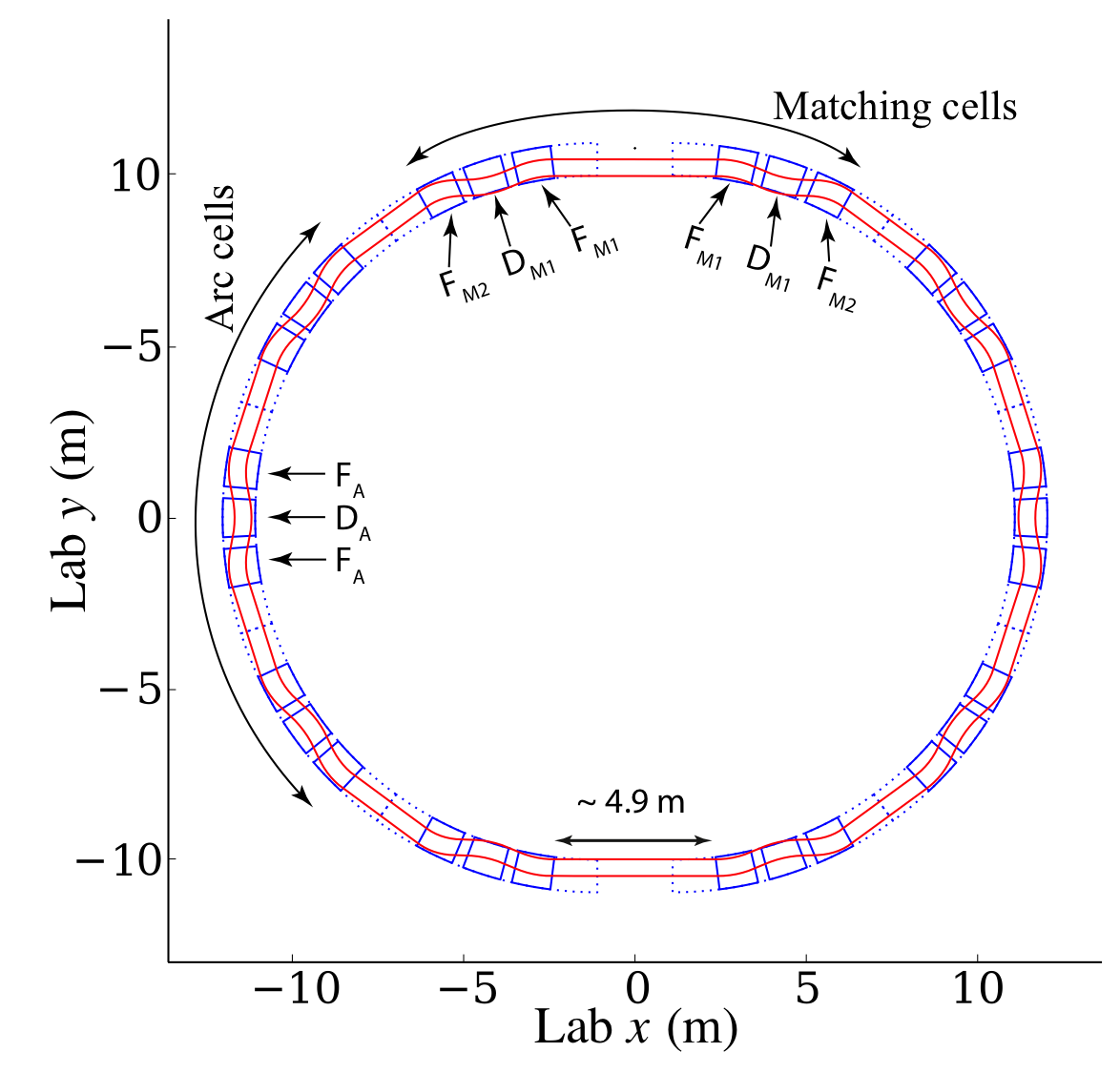}
\caption{Layout of the NORMA racetrack FFAG}
\label{fig:norma2}
\end{figure}

The cell structure (Fig.~\ref{fig:norma1}) in this design was very similar to that of PAMELA, using an FDF triplet. The NORMA study also made further innovations by introducing two long straight sections almost 5~m long for injection, extraction, and acceleration by creating a racetrack-shaped machine, shown in Fig.~\ref{fig:norma2}.

\subsection{Arbitrary-edge-angle non-scaling FFAG designs}
Contemporary with the PAMELA design, an alternative approach by Johnstone and Koscielniak used wedge-shaped quadrupoles to achieve control over the betatron tunes with acceleration~\cite{bib:johnstone07}. In this case both the path length through the quadrupoles and the contribution of edge focusing were designed to be a function of the beam momentum. However, over the energy ranges considered, the orbit shift was relatively large, of the order of 1~m. More information on these designs can be found in Refs.~\cite{bib:johnstone09, bib:johnstone11}.

\section{FFAG gantry designs}
One of the most promising application areas for the FFAG is in the beam delivery system from accelerator to patient. Existing treatment gantries are large, expensive, and slow to provide variable-energy beams to the patient, as one must wait for the magnetic field to be adjusted before a different energy can be accepted through the transport line. In particular, for carbon ions, the existing gantry at the Heidelberg Ion Therapy centre is both large and heavy, weighing in at 630~t, where 135~t of that is the magnets and the remainder is the supporting structure to allow rotation around the patient.

An FFAG gantry would be able to have a large energy acceptance with a single magnet setting and no limit on rapid variation of energy. This means that if the machine could provide energy variability at a rate of 1~kHz, the beam delivery system and gantry would be able to transport this to the patient treatment room.

A transport line and gantry design based on similar optics was developed in the context of the PAMELA project~\cite{bib:machidafenning, bib:fenning12, bib:fenningthesis}. However, in the PAMELA-type design the aperture is still relatively large. Noting that the linear non-scaling FFAG can have an extremely compact aperture over a large momentum range has motivated significant development work on the use of this technique for gantries. Resonance crossing and injection or extraction are obviously not issues in single-pass gantries. The major advantages that could be realized using this technology are reductions in the size, cost, and complexity of treatment gantries.

A design exists for a superconducting non-scaling FFAG gantry for carbon ions~\cite{bib:trbojevic1, bib:trbojevicbnl, bib:trbojevic2} which may reduce the weight of the magnetic components of such a design from 135~t to around 2~t. There are also designs for compact proton FFAG gantries utilizing either compact magnets or even permanent magnets~\cite{bib:trbojevic14, bib:trbojevic3}.

\section{Summary}

I hope you have gained from these notes a basic understanding of FFAG accelerators and their potential for use in medical applications. Here we have focused on hadron therapy applications but, of course, in the high-intensity regime FFAGs also hold promise for other medical applications such as radioisotope production and boron neutron capture therapy. Listed throughout are many references which will provide the reader with a starting point for learning more about the application of FFAG accelerators.

\section*{Acknowledgements}

The author would like to thank members of the FFAG community who have contributed valuable references, images, and discussions. In particular, I would like to thank S. Tygier, F. M\'eot, S. Machida, and D. Kelliher.

\end{document}